# Extended Analysis of Gravitomagnetic Fields in Rotating Superconductors and Superfluids

M. Tajmar[*]

*ARC Seibersdorf research GmbH, A-2444 Seibersdorf, Austria*

C. J. de Matos[†]

*ESA-HQ, European Space Agency, 8-10 rue Mario Nikis, 75015 Paris, France*

**Abstract**

Applying the Ginzburg-Landau theory including frame dragging effects to the case of a rotating superconductor, we were able to express the absolute value of the gravitomagnetic field involved to explain the Cooper pair mass anomaly previously reported by Tate. Although our analysis predicts large gravitomagnetic fields originated by superconductive gyroscopes, those should not affect the measurement of the Earth gravitomagnetic field by the Gravity Probe-B satellite. However, the hypothesis might be well suited to explain a mechanical momentum exchange phenomena reported for superfluid helium. As a possible explanation for those abnormally large gravitomagnetic fields in quantum materials, the reduced speed of light (and gravity) that was found in the case of Bose-Einstein condensates is analysed.

[*] Principal Scientist, Space Propulsion, Phone: +43-50550-3142, Fax: +43-50550-3366, E-mail: martin.tajmar@arcs.ac.at
[†] General Studies Officer, Phone: +33-1.53.69.74.98, Fax: +33-1.53.69.76.51, E-mail: clovis.de.matos@esa.int

**Introduction**

The authors recently published a paper[1], suggesting for the first time, that a reported disagreement between experimental measurements and theoretical predictions for the Cooper pairs mass[2,3] might arise from a non-classical high-order gravitomagnetic contribution (also known as frame dragging or Lense-Thirring effect). In normal matter, the ratio between electromagnetic and gravitational fields is given by the difference in the respective permeabilities[4]. However, magnetic fields generated by rotating superconductors as a consequence of the quantization of the canonical momentum of the Cooper pairs do not depend on the permeability. Hence, there is the possibility that the ratio between those two fields might be different in a quantum material.

Tate et al[2,3] used a SQUID to measure the magnetic field generated by a thin Niobium superconductor, also called the London moment. This experiment was done as a preparation activity for the Gravity-Probe B satellite. As the thickness of the ring was in the order of the London penetration depth, the magnetic field will be zero at certain frequency steps defined by the quantization number $n$ according to the Ginzburg-Landau equation. This zero-flux condition was then used to derive $\hbar/m$. In our previous paper[1], we added a gravitomagnetic component $\Delta B_g$ to this frequency step $\Delta \upsilon$ in the zero-flux condition to account for the deviation of the measured Cooper-pair mass compared to its theoretical value. This gravitomagnetic contribution was found to be $\Delta B_g = 1.65 \times 10^{-5}$ rad.s$^{-1}$ (value corrected[1] from Ref. 1).

In this paper, we will extend our argument and express the absolute value of the gravitomagnetic field possibly involved in rotating superconductors to account for the theoretical disagreement, and it will be applied to further experiments with superfluids, and its possible consequences on the results to be expected from the Gravity-Probe B satellite will be investigated.

---

[1] The corrected equation is $\dfrac{\hbar}{m} = 2S\left(\Delta \upsilon + \dfrac{\Delta B_g}{4\pi}\right)$

## Gravitomagnetic Fields in Superconductors

Tate et al[2,3] followed Ginzburg-Landau theory, integrating the current density of Cooper pairs around a closed path including the effect of a rotating reference frame, but neglecting any gravitomagnetic fields, since the densities and speeds involved are very small from a relativistic point of view,

$$\frac{m^*}{e^2 n_s} \oint_\Gamma \vec{j} \cdot d\vec{l} = \frac{nh}{2e} - \int_{S_\Gamma} \vec{B} \cdot d\vec{S} - \frac{2m^*}{e} \cdot \vec{\omega} \cdot \vec{S}_\Gamma, \quad (1)$$

where $n_s$ is the Cooper-electron number density, $S_\Gamma$ the area bounded by the line $\Gamma$, $\omega$ the angular velocity, $B$ the London moment and $m^*$ the Cooper-pair mass measured by Tate. The argument put forward in Ref. 1 says that Tate's measured Cooper-pair mass $m^*=1.82203 \times 10^{-30}$ kg is different from the theoretically expected one $m=1.82186 \times 10^{-30}$ kg because Equ. (1) does not account for a possibly involved gravitomagnetic field $B_g$. The complete Ginzburg-Landau equation should then read as

$$\frac{m}{e^2 n_s} \oint_\Gamma \vec{j} \cdot d\vec{l} = \frac{nh}{2e} - \int_{S_\Gamma} \vec{B} \cdot d\vec{S} - \frac{m}{e} \int_{S_\Gamma} \vec{B}_g \cdot d\vec{S} - \frac{2m}{e} \cdot \vec{\omega} \cdot \vec{S}_\Gamma. \quad (2)$$

Combining both equations we can express the gravitomagnetic field $B_g$ as

$$\vec{B}_g = 2\vec{\omega} \cdot \left(\frac{m^* - m}{m}\right) + \left(\frac{m^* - m}{m}\right) \cdot \frac{1}{S_\Gamma e n_s} \oint_\Gamma \vec{j} \cdot d\vec{l}. \quad (3)$$

In a superconductor that is thick compared to the London penetration depth, the current integral in the above equations can be set to zero as there is always a path inside the superconductor in which no current is flowing (outside the London penetration depth). The gravitomagnetic field is then just a function of the angular velocity and of the mass difference of the Cooper-pairs

$$\vec{B}_g = 2\vec{\omega} \cdot \left(\frac{m^* - m}{m}\right) = 2\vec{\omega} \frac{\Delta m}{m}. \quad (4)$$

According to Tate's results, the gravitomagnetic field in her configuration but using a thick rotating Niobium superconductor would then be

$$\vec{B}_g = \vec{\omega} \cdot 1.85 \times 10^{-5}, \quad (5)$$

which is very large even for small angular velocities compared to the gravitomagnetic field produced by the Earth (about $10^{-14}$ rad.s$^{-1}$), but can not be ruled out based on the experimental results achieved so far. Unless experimentally proven or disproved,

Equ. (5) is the only possibility to explain the mass anomaly observed by Tate, after reviewing all theoretical approaches[5].

What could cause such abnormally high gravitomagnetic fields? The gravitomagnetic field of a rotating ring can be calculated by:

$$B_g = \frac{\mu_{0g}}{2r} \cdot \frac{dm}{dt} = \frac{4\pi G}{c^2} \cdot \frac{\rho \omega A}{2},\qquad(6)$$

where $\rho$ is the density of the ring, $A$ the cross section, and $c$ the speed of light (and gravity) in the material. Applying Tate's case we arrive classically at

$$\vec{B}_{g,classical} = \vec{\omega} \cdot 3.19 \times 10^{-35},\qquad(7)$$

which is of course well below the result in Equ. (5). The abnormal coupling could then be due a different value for $G$ or $c$ in the case of quantum materials. If Newton's gravitational constant would be drastically different for superconductors, then the superconductor should change its weight passing through its transition temperature. According to recent measurements[6], this is not the case within an experimental accuracy of 0.00014% for YBCO.

However, a huge index of refraction and therefore an extremely low speed of light was indeed noted in recent experiments with Bose-Einstein-Condensates[7]. Comparing Equs. (5) and (6), a speed of light of $3.9 \times 10^{-7}$ m.s$^{-1}$ would be needed to match our conjecture and Tate's Cooper pair mass anomaly. Liu et al were able to first reduce the speed of light by seven orders of magnitude and then in a second experiment to apparently stop light in a Bose-Einstein-Condensate made out of sodium atoms. Such behaviour could give an explanation why rotating quantum materials could produce non-classical gravitomagnetic fields – as outlined in our conjecture to find a solution to Tate's results.

Chiao recently carried out an experiment to test if a quantum fluid can act as a transducer between gravitational and electromagnetic waves[8]. He performed a Hertz-type measurement and was not able to measure any interaction between both fields. According to the resolution of his experiment, he gave an upper limit for the conversion efficiency of $\eta < 1.6 \times 10^{-5}$ (if one assumes equal conversion efficiency from one form of wave to another). This is actually close to our "conversion efficiency" between angular velocity and gravitomagnetic fields based on Tate's measurements in Equ. (5). However, in Chiao's case, the superconductor was not rotating but subjected to EM fields. Therefore, only the Cooper-pairs will move to produce a current, but the bulk of the superconductor is at rest. Following our arguments above, only the Cooper-pair density would then contribute to the generation of gravitomagnetic fields according to Equ. (6). For the case of Niobium similar to Tate, this density is 5 orders of magnitude below the bulk density of Niobium. Therefore, gravitomagnetic fields generated by external EM fields would be correspondingly 5 orders of magnitude smaller than those generated by non-stationary superconductors. Taking into account the different cross sections for Chiao's and Tate's measurements (only the area within the London penetration depth contributes), we would then expect a "conversion efficiency" of $\eta \cong 10^{-7}$, well below Chiao's upper limit.

In conclusion, we expect a Faraday-type experiment to be more efficient compared to a Hertz-type experiment with respect to the generation of gravitational waves.

NASA and Stanford University launched the Gravity-Probe B (GP-B) satellite in April 2004 to measure the gravitomagnetic field (Lense-Thirring effect) of the Earth[9] by precession of superconductive gyroscopes relative to a fixed star. The gyroscope consists of a fused quartz ball covered with a 1.5 µm thick Niobium layer. This ball is rotated at a temperature of 2 K and therefore produces a magnetic field being a rotating superconductor. The change of axis orientation is measured by SQUID magnetometers. According to general relativity, the geodetic precession rate is 6.6 arc-sec/yr and the Lense-Thirring precession rate is 0.042 arc-sec/yr . As these gyroscopes are rotating superconductors with a Niobium thickness larger than the London penetration depth (about 85 nm), a gravitomagnetic field according to Equ. (5) should be generated if our conjecture is correct. Would that influence the measurements on GP-B?

The spin angular momentum of GP-B's superconductive gyroscopes, measured by a co-moving observer, $\vec{S}_0$, obeys the equation

$$\frac{d\vec{S}_0}{dt} = \vec{\Omega} \times \vec{S}_0 , \tag{8}$$

where $\vec{\Omega}$ includes the Thomas-precession, the geodetic and the gravitomagnetic (Lense and Thirring frame dragging) precessions[10].

Our conjecture leads to a modification of the superconductive gyroscope's spin angular momentum according to:

$$\vec{S}_0 = I_{SC}\omega_{SC}\left(1 + 2\frac{\Delta m}{m}\right), \tag{9}$$

where $I_{SC}$ and $\omega_{SC}$ are the moment of inertia and the angular momentum of the superconductive gyroscopes respectively. The product $S_{c0} = I_{SC}\omega_{SC}$ is the classical angular momentum measured in the co-moving frame. Doing (9) into (8) leads to:

$$\frac{d\vec{S}_{c0}}{dt} = \vec{\Omega} \times \vec{S}_{c0} . \tag{10}$$

The quantum contribution to the angular momentum cancels out, leaving the usual relativistic equation, ruling the behaviour of GP-B's gyroscopes, unchanged. Therefore the anomalous gravitomagnetic moment predicted by our conjecture will have no influence on the geodetic and gravitomagnetic precessions measured in the GP-B experiment!

## Gravitomagnetic Fields in Superfluids

In our previous paper[1], we discussed the quantization of the full canonical momentum including a gravitomagnetic vector potential $A_g$ for the case of a neutral superfluid:

$$\oint \vec{p}_s \cdot d\vec{l} = \oint \left( m\vec{v}_s + m\vec{A}_g \right) \cdot d\vec{l} = nh. \tag{11}$$

As there is no comparible situation like a thick superconductor where $n$ can be set to zero, the gravitomagnetic field generated by a superfluid can not be expressed easily as for thick superconductors in Equ. (4). However, according to the Feynman-Onsager-vortex model, the first vortex in a superfluid ($n$=1) only appears if the angular velocity is larger than a critical angular velocity[11] $\omega_c$

$$\omega_c = \frac{\hbar}{mR^2} \ln\left(\frac{R}{a}\right), \tag{12}$$

where $R$ is the inner radius of the superfluid cylindrical container and $a$ is the effective radius of the vortex core. In the case of superfluid helium-I, $a \simeq 10^{-10}$ m. For a container's rotational speed below $\omega_c$ the superfluid will be in a non-rotational state, which is also called the rotational Meissner effect. This has been experimentally confirmed by Hess and Fairbank[11]. In this specific case, we can write Equ. (11) as

$$\vec{B}_g = -2\vec{\omega} \quad \omega \leq \omega_c. \tag{13}$$

But due to the non-rotational state of the superfluid, the angular velocity of the superfluid will be zero and hence $B_g$=0. Let us now switch reference frames between the superfluid bulk and the container, this is equivalent to the case that the container is at rest and the superfluid is rotating. Here Equ. (13) should apply and an equivalent gravitomagnetic field produced accordingly. For that case, Equ. (13) would correspond exactly to the gravitational analogue of the magnetic Barnett effect, that we formerly called the "Gravitomagnetic Barnett Effect"[12].

For angular velocities of the container greater than $\omega_c$, a gravitomagnetic field could be generated, modulated by a function similar to Equ. (3) in the case of thin superconductors, where we can write

$$\vec{B}_g = \frac{n}{S}\frac{h}{m} - 2\vec{\omega} \quad \omega > \omega_c, \tag{14}$$

where $n$ is the total number of vortices present in the superfluid and $S$ is the integration area. The conservation of angular momentum in rotating superfluids would therefore require the generation of a gravitomagnetic field through the cross section of the rotating container. In the case of multiply-connected containers, this field should also be present in the volume that does not contain the superfluid; similarly as for the magnetic case with rotating superconductors.

Due to accelerations and decelerations of the container, the changing gravitomagnetic fields will generate so-called gravitoelectric (also called non-Newtonian gravitational) fields[4], which would then act on the superfluid, as it has mass:

$$rot\ \vec{g} = -\frac{d\vec{B}_g}{dt}.  \tag{15}$$

In fact this seems to be a logical way to add momentum to the fluid in its superfluid state as there is no friction and no charge (besides using large magnetic fields that act on the small magnetic moments of the helium atoms). Due to the double minus sign in Equs. (13) and (15), the gravitational field would act in the same direction as $\omega$.

This behaviour was actually measured in a follow up paper from Hess after their detection of the rotational Meissner effect in superfluids[13]. He mounted the superfluid helium container on an extremely low friction magnetic bearing to study the momentum exchange of the superfluid with the container. After cooling the helium at rest below its critical temperature, he rotated the container and observed that nearly all momentum that he injected in the container was transferred to the superfluid (vorticies appeared) – although a superfluid has no friction. Also the other direction worked in a similar way: the liquid was first cooled during rotation below its critical temperature. Then the angular momentum of the container was reduced and, accordingly, also the superfluid was found to loose vortices and momentum. Hess called it mechanical generation/deletion of vortices in superfluids – without an apparent explanation. According to the knowledge of the authors, no subsequent experiments of a similar kind have been carried out to further characterize this phenomenon in superfluids.

We believe, that this superfluid behaviour adds strong support to the conjectural large gravitomagnetic fields as outlined in this paper.

**Conclusions**

We extended the analysis of our previous paper to show how large gravitomagnetic fields could be involved in rotating superconductors and superfluids. An absolute value of the gravitomagnetic field necessary to explain Tate's Cooper pair mass anomaly was derived. According to our further analysis, an explanation for this abnormally large gravitomagnetic field could be the much lower speed of light (and gravity) measured for Bose-Einstein condensates. Comparing this theoretical approach with a recent measurement on the conversion efficiency between electromagnetic and gravitational waves, we find that Faraday-type induction experiments are orders of magnitude more efficient than Chiao's Hertz-type induction experiment. The conversion efficiency predicted by our theoretical approach is well below Chiao's measured upper limit.

Although being a large effect compared to the gravitomagnetic field generated by the Earth itself, the influence on the measurements to be performed by the Gravity-Probe B satellite appears to be irrelevant.

After further analysing the case of rotating superfluids, we found also strong support for our conjecture being able to explain the mechanical generation/deletion of vortices in superfluid helium by rotation of the superfluid's container. This is an intriguing simple argument to explain frictionless momentum exchange between the superfluid and the container.

These results shall stimulate experimental and theoretical work to prove or disprove the conjectured large gravitomagnetic fields postulated. According to our analysis, the fields should be large enough to be detected in a laboratory setup investigating non-stationary rotational regimes. Presently, an experiment is underway at ARC Seibersdorf research to test the conjecture outlined in this paper.